\documentstyle[prl,aps,multicol,epsfig]{revtex}
\begin{document}
\draft
\widetext
\title{ Quantum critical point with competing propagating and
diffusive spin excitations
 }
\author{J\"org Schmalian    }
\address{ University of Illinois at 
Urbana-Champaign,   Loomis Laboratory 
of Physics, 1110 W. Green,  Urbana, IL, 61801 }
\date{ \today}
\maketitle
\widetext 
\begin{abstract} 
\leftskip 54.8pt
\rightskip 54.8pt
Feedback effects due to spin fluctuation induced   precursors 
in the   fermionic quasiparticle spectrum are taken into
account in the description of a quantum critical point of itinerant
spin systems. A correlation length dependent spin damping
 occurs,
  leading to a   dynamical scaling  
  with $z\approx 1$ which non-trivially
competes  with the conventional spin wave behavior.
 We obtain, within a one loop renormalization
group approach, a quantitative  explanation for the scaling behavior seen 
 in   
underdoped cuprate superconductors.
\end{abstract}   
\begin{multicols}{2}
\narrowtext   
Nuclear magnetic resonance
(NMR) and inelastic neutron scattering (INS) experiments in magnetically
underdoped cuprates~\cite{BP95,ZBP96,AMH97,CCS96} indicate that the  spin excitations of these
systems  are close to a
zero temperature quantum critical 
point (QCP)\cite{Hertz,CHN88,Millis93,CS93,SCS95}.
These     measurements   give strong
indications for a dynamical scaling exponent $z\approx 1$, 
typical for systems characterized by propagating 
antiferromagnetic (AF) spin modes,
but different from that    expected for    an itinerant
 AF system with $z=2$\cite{Hertz,Millis93,SCS95}.
On the other hand, 
the low frequency  behavior  of the imaginary  part of the dynamical
spin susceptibility is characterized 
by over-damped spin-excitations, i.e. ${\rm Im}\chi({\bf q},\omega)
\propto \omega +{\cal O}(\omega^3) $, as expected for itinerant
 AF.
Thus,   while the  spin dynamics is clearly
over-damped, the scaling  behavior  corresponds to that of a  system 
without damping,   causing a conceptual problem for the description of the 
related QCP.

In the present paper we investigate the behavior in the vicinity 
of a  QCP  characterized by  competing propagating and
diffusive spin excitations. 
We explicitly take   into account   
that for a system characterized by    strong interactions between 
  collective
spin modes and    fermionic quasiparticles 
 the latter are strongly affected by
the critical mode. This, in turn,  leads     to a feedback in the
 spin dynamics~\cite{MPchem,SPS98}, causing $z\approx  1$
   dynamical scaling behavior for an  itinerant AF.  
Our theory offers a quantitative explanation for the temperature
and frequency  dependence of INS and NMR data and 
confirms     
 the phenomenological description of the spin dynamics
of Refs.~\cite{BP95,ZBP96}.

\begin{figure}
\vskip -0.2 cm
\centerline{\epsfig{file=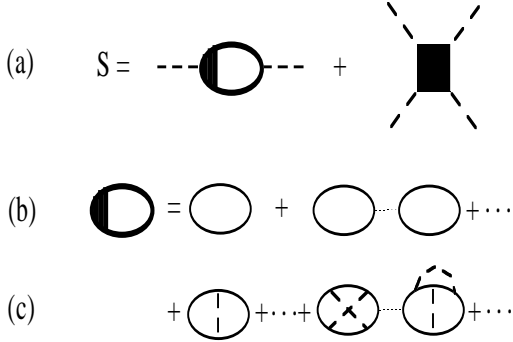,width=7cm, height=4.5cm,
    scale=0.85}}
\vskip 0.2cm
\caption{(a) Diagrammatic representation of the effective action with
   spin-propagator and spin spin interaction term. The
    dashed lines correspond to collective spin degrees, the solid
  lines to  the renormalized fermionic Green's function.
(b) Typical weak coupling approximation for $\chi_0(q)$. 
Dotted
lines  refer to the fermion-fermion interaction, $f_{\bf q}$.
(c) Some of the   diagrams causing  
feedback effects of the quasiparticle excitations.}
 \label{fig1}
\end{figure}

We start from an interacting Fermi system characterized by some
unperturbed band-part and  an   interaction term, 
$\sum_{{\bf q}} f_{\bf q} 
\, {\bf s}_{\bf q} \cdot {\bf s}_{-{\bf q}} $,
with  fermionic 
spins,   ${\bf s}_{\bf q} =\frac{1}{2}\sum_{{\bf k} \sigma \sigma'} 
c^\dagger_{{\bf k}+{\bf q} \sigma} 
{\bf \sigma}_{\sigma \sigma'} c_{{\bf k} \sigma'}$. 
By introducing a collective spin-1 Bose field, ${\bf S}(q)$,  via  the 
Hubbard-Stratonovich
transformation, one can integrate out  the fermions
and expand  up to  forth
order in ${\bf S}(q)$~\cite{2kf};  
   the  resulting  effective action of the collective spin
degrees of freedom is~\cite{Hertz,Millis93}:
\begin{eqnarray}
S&=&\frac{1}{2} \int^{\Lambda} dq \, \,\chi^{-1}_0(q) \, \,{\bf S}(q) \cdot {\bf S}(-q)
\nonumber \\
& &+u \, \int^{\Lambda} dq_1 \cdots   \int^{\Lambda} dq_4 \, \,
\delta_{q_1+q_2+q_3+q_4}\nonumber \\
& &\ \ \ \ \ 
 \times
 {\bf S}(q_1)\cdot {\bf S}(q_2) \, {\bf S}(q_3)\cdot {\bf S}(q_4)\, ,
 \label{action}
\end{eqnarray}
characterized by a    bare
 spin propagator, $\chi_0(q)$, and coupling constant, $u>0$.
 In Eq.~\ref{action}
the $(d+1)$-dimensional vector $q=({\bf q},i\omega_n)$ consists 
of the $d$-dimensional momentum vector ${\bf q}$ 
and the bosonic Matsubara frequency
$\omega_n=2n\pi T$  at  temperature, $T$.
 We write
$\int^\Lambda dq \cdots = T\sum_n
\int_{|{\bf q}|<\Lambda}\frac{d^d{\bf q}}{(2\pi)^d } \cdots$ and
$\delta_{q+q'}=T^{-1} \delta_{n+n'} \delta^{(d)}({\bf q}+{\bf q}')$, where
$\Lambda$ is the upper momentum cut off.
A diagrammatic representation of this action is given in Fig.~\ref{fig1}(a).

Usually,
  $\chi_0(q)$ and   $u$ are
calculated within a weak coupling approach in which 
the spin    damping due to particle hole
excitations  is identical to that of a noninteracting electron gas~\cite{Hertz,Millis93}
(see Fig.~\ref{fig1}(b)).
 It is  however to be expected  that  the fermionic quasiparticles
are 
strongly affected by their own spin excitations once the system gets 
close to a magnetic 
instability~\cite{                                                                                                                                                                                                                                                                                                                                                                                                                                                                                                                                                                                                                                                                                                                                                                                                                                                                                                                                                                                                                                                                                                                                                                                                                                                                                                                                                                                                                                                                                                                                                                                                                                                                                                                                                                                                                                                                                                                                                                                                                                                                                         SPS98,SDW,Schrieffer,VT97}.
It is therefore  interesting  to take into account these
modifications of the quasiparticle spectrum   in an 
 investigation of the critical
behavior. 
Of particular importance  are      quasiparticles close
to the magnetic BZ-boundary,   which, on the one hand, determine
  the spin damping,
$\gamma \propto {\rm
  Im}\chi^{-1}({\bf Q},\omega)/\omega |_{\omega \rightarrow 0}$, 
   and, on the other hand,
are mostly affected by the proximity to  the ordered state.

 Magnetic precursors in the quasiparticle spectrum
 have recently been investigated in 
several   theories~\cite{                                                                                                                                                                                                                                                                                                                                                                                                                                                                                                                                                                                                                                                                                                                                                                                                                                                                                                                                                                                                                                                                                                                                                                                                                                                                                                                                                                                                                                                                                                                                                                                                                                                                                                                                                                                                                                                                                                                                                                                                                                                                                         SPS98,SDW,Schrieffer,VT97}.
Within a quasistatic theory~\cite{SPS98},  
applicable at higher $T$, $\gamma$ was  
calculated to all orders in the perturbation theory.
Due to precursors of a spin density wave (SDW) 
gap the low energy spectral weight is reduced and the
vertex function enhanced and
 $\gamma$ was    found to 
change  from being a constant to a  $\xi$-dependent
function:  $\gamma \propto
 \xi^{-\varphi} $ with $ \varphi \approx 1$, 
 once     $\xi $ is larger  than the electronic length scale,
  $\xi_0 = 2 v/f_{\bf Q}$.    
In cuprates $\xi_0\approx 2$ due to the large coupling 
constant $f_{\bf Q}$.
For low temperatures the situation is more 
complicated~\cite{comment,CS98}.
Nevertheless, the leading contribution to the spin
 damping  can be estimated as
 $\gamma \propto  f_{\bf Q}^2 \Gamma^{{\rm sf}} /v^2$,
with spin fermion vertex function at the Fermi surface (FS)
and for momentum transfer
 ${\bf Q}\approx (\pi,\pi)$, $\Gamma^{{\rm sf}}$,
and Fermi velocity, $v$. 
At the critical point the vertex function for scattering processes
  including a Goldstone mode is expected to vanish
on a surface defined by momentum and energy conservation
of the involved quasiparticles~\cite{Adler}.
 In the case of   large correlation length and 
large interaction $f_{\bf Q}$,
  a  FS evolution towards an  SDW like
 behavior  occurs and the above principle causes
 the  vertex function to vanish like
$\Gamma_{{\rm sf}} \propto \xi^{-1}$~\cite{Schrieffer}, leading to 
$\gamma \propto  \xi^{-1}$. 
Thus,  we expect in various regimes    
a $\xi$-dependent reduction of the spin damping  
 due to the proximity 
to an ordered state.  

In the present communication we
     take  these  feedback effects  into account and determine $\xi(T)$
using a
renormalization group approach. The propagator 
of the   collective spin modes 
is assumed to be:
\begin{equation}
\chi_0({\bf q},i\omega_n)=\frac{\alpha}{
\xi_0^{-2} + {\bf q}^2 +
 \xi_0^{-\varphi} |\omega_n|/\hat{c}+(\omega_n/c)^2}\, ,
\label{chi}
\end{equation}
with bare
 correlation length $\xi_0$, spin damping, 
$\gamma= \hat{c}^{-1}\xi_0^{-\varphi}$ and spin wave velocity $c$.
Momenta are measured relative to the ordering vector.
 For  $\hat{c}\rightarrow \infty $, i.e. without spin damping,  the problem is similar
to that    investigated by Chakravarty {\em et al.}~\cite{CHN88}  and
Chubukov {\em et al.}~\cite{CS93}.  
The alternative limit  $\varphi=0$ and $c \rightarrow \infty$   was
discussed   by Hertz~\cite{Hertz} and Millis~\cite{Millis93}.
The case $\varphi=0$ but $c $  finite was discussed by
 Sachdev {\em et al.}~\cite{SCS95}, who   found 
a $z=1$ to $z=2$ crossover for decreasing temperature,
in contrast to the experimental finding 
 in  the  cuprates  of  a $z=2$ to $z=1$
crossover for decreasing $T$~\cite{BP95,ZBP96,CCS96}.

We  use   a
  Kadanoff-Wilson 
momentum shell renormalization group approach;  integrating  out states with 
momenta between 
$\Lambda e^{-l}$ and $\Lambda$,  including  the rescaling
   $T(l) = e^{zl} T$  to 
reach  self-similarity.  
Up to one loop the resulting flow equations are:
\begin{eqnarray}
\frac{dT(l)}{dl}&=&z T(l) \label{fl1} \\
\frac{d\xi_0^{-2}(l)}{dl}&=& 2 \xi_0^{-2}(l)+4(N+2) u  \Phi(\xi,T(l))
 \label{fl2} \\
%
%
  \frac{du(l)}{dl}  &=& \epsilon  u(l)  \label{fl3} 
-4(N+8) u^2  \Psi(\xi,T(l))    \\
%
%
\frac{d\hat{c}(l)}{dl}&=&(z+\varphi-2)  \hat{c}(l)  \label{fl4} \\
\frac{dc(l)}{dl}&=&(z -1)c(l) \label{fl5}  \, ,
\end{eqnarray}
with $\epsilon=4-(d+z)$. $N$ is the number of vector
components of   ${\bf S}(q)$.
 The functions $\Phi$ and
$\Psi$ can be 
  determined along the same lines as in 
Ref.~\cite{Millis93}. Their explicit dependence  
on the  renormalized correlation length, $\xi$,
 results from  a replacement of 
$\xi_0$ by $\xi$ in  diagrams like those in Fig.~\ref{fig1}(c)
which determine the    spin damping,
  necessary to reach self
consistency~\cite{Subir94}.
 From Eqn.~\ref{fl4}
 and \ref{fl5} we find that for $\varphi=1$ (i.e. $z=1$) the   velocities
 $\hat{c} $ and   $c$ do not renormalize in the one loop approximation.
In the case 
  $\varphi < 1$  one 
starts
from a primary scaling behavior with $z=1$  which  
eventually crosses over to $z=2-\varphi $\cite{SCS95}. 
 In what follows  we assume    
$\varphi=1$    since quantitatively the
 latter crossover will barely change our results; an extension
to arbitrary $\varphi$ is   straightforward. 
Moreover, small
deviations from $z=1$ seem to be beyond  the accuracy of the
current  experiments.
 For $\epsilon >0$
the  flow equations, Eqn.~\ref{fl1} - \ref{fl5}, are characterized by a zero temperature
fixed point
 $(\xi^*_0)^{-2}  =-\frac{N-2}{N-8}  \epsilon \Lambda^2$ and 
$u^*=\Gamma(d/2) 2^{d-1} \pi^{d/2} \epsilon/((N+8)c 
\Lambda^{d-3}$, which is 
unstable since $T$ is a relevant field.
By integrating the flow equations we find that 
  the critical behavior is sufficiently characterized 
by the single effective coupling constant 
$g= 2(1-d)(N+2) u \xi_0^2$.
This is also expected since the critical behavior should be similar to that of
the   quantum nonlinear sigma model which is  fully characterized by
$g$ and $T$~\cite{CHN88,CS93}.
For $d=2$,  the   fixed point value of $g$ is given by
$g^*=4 \pi/(c \Lambda)$, which
   is   unchanged  from the
result without damping~\cite{CHN88,CS93}
 (the limit $\hat{c} \rightarrow \infty$).
Nevertheless, the critical
behavior in the vicinity of this QCP depends strongly
on the ratio, $\Gamma \equiv c/\hat{c}$.
In the limit $c\rightarrow \infty$    the QCP moves towards 
$g^*=8\pi/\Lambda_\omega$ with upper frequency 
cut off $\Lambda_\omega$.
The flow equations, Eqn.~\ref{fl1} - \ref{fl5}, can be integrated and the correlation
length follows from $\xi={\rm e}^l \xi(g(l),T(l))$, where we determine
 the
correlation length at  the 
matching point~\cite{CHN88}, $l=\bar{l}$, with 
$T(\bar{l})=c\Lambda/4$ and $g(\bar{l})=2g^*$  from 
 a $1/N$ expansion as used in Ref.~\cite{CS93}.
Note, since   $g(l)$ depends   on $\xi$, this
procedure leads to a self consistency condition for $\xi$.

\begin{figure}
\vskip -0.2 cm
\centerline{\epsfig{file=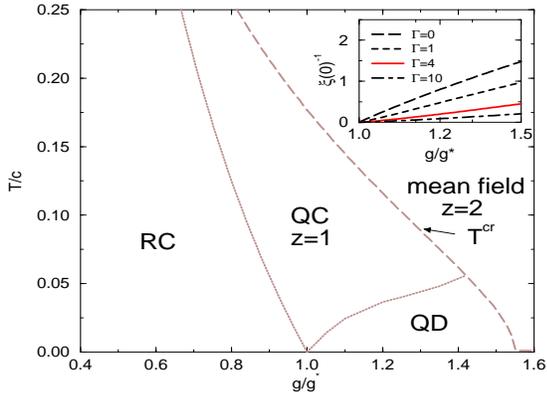,width=8cm, height=6cm,
    scale=0.85}}
\vskip 0.2cm
\caption{Phase diagram of an itinerant  AF
with $z=1$ scaling behavior for $\Gamma= 4$.
 RC is the renormalized 
classical regime with exponentially large  $\xi$.
  QC is the quantum critical regime
and QD the quantum disordered regime. In the mean field
regime, $T>T^{cr}$, with $\xi<\xi_o\approx 2$ no feedback effect due
to changed quasiparticle behavior occurs, leading to
   $z=2$. The inset shows the
inverse zero temperature correlation length in the QD regime for
different  $\Gamma$.}
 \label{fig2}
\end{figure}

The QCP,  $g^*$, separates in the usual sense~\cite{CHN88}
a renormalized classical (RC) regime 
with  
exponentially growing correlation length,
$\xi(T)=\hat{c}/(2T) \exp{(2\pi c
 \Lambda/T(\frac{1}{g^*}-\frac{1}{g}))}$,
for $g(l=0)<g^*$, from
a quantum disordered (QD)   state with 
 a finite zero temperature correlation 
length, $\xi(0)$, (see Fig.~\ref{fig2}).
 We find for  $\Gamma \gg1$ that 
 $\xi(0)/\log(\xi(0) \Lambda) =\frac{\Gamma g^*}{4\pi \Lambda} 
 (g-g^*)^{-1}$, which is
     enhanced  compared to the situation without damping:
$\xi(0)= \frac{g^*}{2 \Lambda} (g-g^*)^{-\nu}$
with $\nu=1$ up to one loop.
 The results of a    numerical evaluation of $\xi(0)^{-1}$     are 
  shown in the  inset of Fig.~\ref{fig2}.

\begin{figure}
\vskip -0.2 cm
\centerline{\epsfig{file=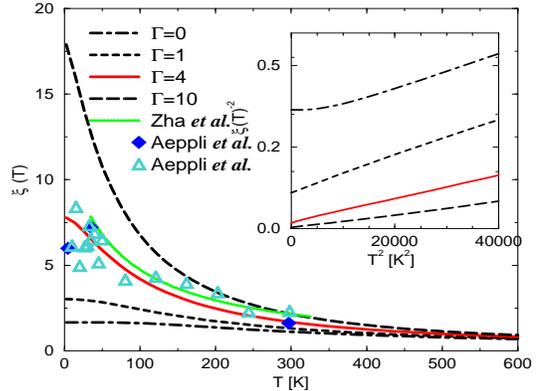,width=8cm, height=6cm,
    scale=0.85}}
\vskip 0.2cm
\caption{ Temperature dependence of  the correlation length 
for $g=1.18g^*$ and  different $\Gamma$  
in   comparison with experimental data for
 La$_{1.86}$Sr$_{0.14}$CuO$_4$
from Ref.[3] at $\omega=2.5\, (3.5)\, {\rm meV}$ indicated by 
diamonds (triangles) and the phenomenological  theory of Ref.[2]. 
 The inset shows $\xi(T)^{-2}$ versus $T^2$.}
 \label{fig3}
\end{figure}

Modified  scaling behavior    can 
in similar fashion be found 
 as function of  temperature.
 In Fig.~\ref{fig3} we show our results for the $T$-dependent correlation 
 length for $g/g^*=1.18$  and
 different values of $\Gamma$. We find
 up to logarithmic corrections,  
$\xi(T)^{-2} \approx  \xi(0)^{-2}+bT^2$ (see inset of Fig.~\ref{fig3}(b)).
 In contrast, in  the case of propagating spin excitations a  sharp
 transition  to an expnentially weak   temperature 
dependent $\xi$ occurs for 
 $T\approx \Delta$ with spin wave gap
 $\Delta =c\xi(0)^{-1}$.  Thus, due to spin damping 
  the spin wave gap is   filled with 
 low energy states and the correlation length  continues to grow 
 even  for  very  low temperatures.
The crossover between the quantum critical  (QC) and QD 
regime is very gradual.
In Fig.~\ref{fig2} the crossover line was determined by the temperature
where $d^2\xi^{-1}/dT^2 $ becomes small compared to its low $T$ value,
i.e. where $\xi^{-1} $ starts  to grow linearly with $T$.
Finally, considering  higher temperatures,  we have to take into account
 that  $\gamma \propto \xi^{-1}$
occurs only   for $\xi \geq \xi_o\approx 2 $~\cite{SPS98}.  
For $\xi<\xi_o$, $\gamma =const.$  leading to 
a $z=2$ behavior at a characteristic temperature $T^{\rm cr}$ 
(see Fig.~\ref{fig2}).

 Another characteristic 
   phenomenon caused by the proximity to a QCP 
   is   the scaling behavior
of the  frequency dependence of ${\rm Im}\chi({\bf q},\omega)$.
For any point in the $(T,g)$ phase diagram 
(except
 $g\leq g^*$ and $T=0$)
 ${\rm Im}\chi_{\bf q}(\omega)$
 increases linearly  with  $\omega$  due to excitations
 in the particle hole continuum.
In the RC regime however, the spin damping is exponentially suppressed and
the spin dynamics is indistinguishable from a system with purely propagating
spin waves.
  On the other hand, in the  QC regime  $\omega/T$ scaling behavior
 of   $\xi^{-2} {\rm Im}\chi_{{\bf q}=0}(\omega)$ and of
 the momentum averaged susceptibility $\int d^2{\bf q} {\rm Im}\chi_{\bf q}(\omega)$
 is found and  the low frequency slope of $\chi_{{\bf q}=0}(\omega) $
 behaves like
$ {\rm Im}\chi_{{\bf q}=0}(\omega)/\omega |_{\omega \rightarrow 0} 
 =\alpha  \xi^{\delta}/\hat{c}$ with $\delta=3$.

\begin{figure}
\vskip -0.2 cm
\centerline{\epsfig{file=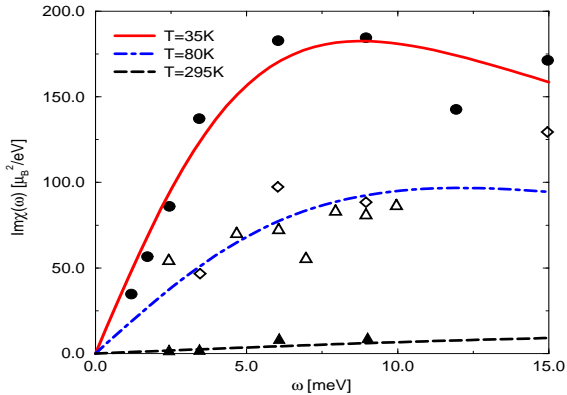,width=8cm, height=6cm,
    scale=0.85}}
\vskip 0.2cm
\caption{ Frequency dependence of the dynamical 
 susceptibility at the
  peak maximum for $\Gamma=4$ and $g=1.18g^*$
in comparison with INS experiments from Ref.[3].}
 \label{fig4}
\end{figure}

 In  applying    these results   to cuprate 
 superconductors,  we  assume that   doped systems without
 long range order are located in the QC and  QD regime.
 Since $g\propto 1/\langle {\bf S}({\bf r})^2\rangle$,  as follows from a $1/N$ expansion,
 doping  reduces 
   the effective moment, causing the coupling constant to grow 
 until it exceeds $g^*$. 
  For the    cuprate material La$_{1.86}$Sr$_{0.14}$CuO$_4$
  the parameters $\hat{c}\approx 50 \, {\rm meV}$
  and $c\approx  220 {\rm meV}$ are reasonably 
  well
  known from NMR  and  INS experiments~\cite{BP95,ZBP96,AMH97} and  it follows
  $\Gamma \approx 4$.
Thus,   it     suffices for a quantitative  understanding of the INS data of
  Ref~\cite{AMH97}  to  determine  the ratio $g/g^*$ from the correlation
  length at a given 
  temperature. Once this     number is determined 
  one has a complete description  of the  $T$- and $\omega$-dependence
of the  spin fluctuation spectrum.
 As shown in Fig.~\ref{fig3}, a value $g=1.18g^*$
 gives a reasonable agreement not only for the value
of $\xi$ at  low $T$ but  also for the whole temperature regime 
up to $300\, {\rm K}$.
Using this value for the coupling constant we can 
determine the frequency and temperature dependence 
of the dynamical spin susceptibility. 
The results for ${\rm Im}\chi_{{\bf q}=0}(\omega)$ with $\alpha=26\, {\rm eV}^{-1}$,
  shown in Fig.~\ref{fig4}, 
are in remarkable agreement with the  results  of Aeppli {\em et al.}~\cite{AMH97},  who also find  
$ {\rm Im}\chi_{{\bf q}=0}(\omega)/\omega |_{\omega \rightarrow 0} \propto
\xi^{\delta}$ with $\delta=3\pm 0.3$.
This   agreement between theory and experiment 
is a direct consequence of the  fact that the experimental data  show a $\omega/T$
scaling behavior of $\xi^{-2} {\rm Im}\chi_{{\bf q}=0}(\omega)$ 
which is a strong indication that
the system under consideration is indeed close to a 
QCP. 
In order to demonstrate that our results agree quantitatively with the 
phenomenological description
of the $z=1$ pseudo-scaling  regime  of Ref.~\cite{ZBP96}, we also 
show in Fig.~\ref{fig3} the correlation length, $\xi(T)$, as 
  obtained from NMR experiments~\cite{ohg94}.

In conclusion, 
 based on previous 
calculations~\cite{                                                                                                                                                                                                                                                                                                                                                                                                                                                                                                                                                                                                                                                                                                                                                                                                                                                                                                                                                                                                                                                                                                                                                                                                                                                                                                                                                                                                                                                                                                                                                                                                                                                                                                                                                                                                                                                                                                                                                                                                                                                                                         SPS98,SDW,Schrieffer,VT97},
we have  argued that  modifications
of the particle hole excitation spectrum 
 can change the dynamical
scaling behavior of itinerant  AF systems close
to a QCP by affecting the spin damping.
 The resulting $z\approx 1$ scaling causes a 
 T-dependence of the  AF correlation length which
is completely different from  the usual $z=2$ case. 
This new $T$-dependence  
of the dynamical spin susceptibility  is in remarkable agreement with  
  NMR and INS  experiments  and gives an 
explanation for the crossover
scenario  of Refs.~\cite{BP95,ZBP96}.  
 The  $z =1$   scaling and the position of the 
  QCP are  the same  as those for propagating spin modes
in insulating  AF. However, due to damping, 
 the   correlation length is enhanced 
  and the $T$-dependence of $\xi$  
 is   changed.

 We expect   similar behavior in  the
  quantum phase transitions of  other itinerant systems   if the 
 strong interaction between quasiparticles and   collective modes
    changes the   dynamics 
of the collective mode under consideration.
Examples are one and two dimensional charge density 
wave systems, superconductors
or saturated ferromagnets.
In higher dimensions   or for systems with 
weak quasiparticle-collective mode coupling
precursor phenomena are only expected once 
$\xi$ diverges~\cite{VT97}
 and will  
play no role in the quantum disordered regime.

Finally,  we note 
 that the theory presented in this paper is 
  not a theory for the crossover to a  
{\em strong pseudogap } state     found in   
  many cuprate superconductors at low $T$, where
  a decoupling of the $T$-dependence of $\xi$ and  
$\gamma$ has been  found~\cite{BP95,CCS96}. This   might be  
 due to the interference  with   excitations 
different from the   spin fluctuation channel 
discussed here~\cite{CS98,Grilli}.
Nevertheless, we expect   the proximity to  a QCP       to be
essential for  the appearance of the strong pseudogap  behavior.

This work has been supported in part by the Science and Technology
Center for Superconductivity through NSF-grant DMR91-20000
and by the Deutsche Forschungsgemeinschaft.
I thank  A. V. Chubukov, M. Grilli, D. Morr,  D. Pines,  R.
 Ramazashvili, D. Scalapino, and
B. Stojkovi\'c for helpful discussions.

\end{multicols}

\begin{thebibliography}{99}
\bibitem{BP95} V. Barzykin and D. Pines, Phys. Rev. B {\bf 52}, 13585
(1995).
\bibitem{ZBP96} Y. Zha, V. Barzykin, and D. Pines,
                         Rev. B {\bf  54}, 2561 (1996).
%
\bibitem{AMH97} G. Aeppli, T. E. Mason, S. M. Hayden, H. A. Mook, J. 
                           Kulda, Science {\bf 278}, 1432 (1997).
\bibitem{CCS96}N. J.  Curro, T. Imai, C. P. Slichter,
and B. Dabrowski, Phys. Rev. B {\bf 56}, 877 (1997).
%
\bibitem{Hertz} J. A. Hertz, Phys. Rev. B {\bf 14}, 1165 (1976).
%
%
\bibitem{CHN88} S. Chakravarty, B. I. Halperin, and D. R. Nelson,
Phys. Rev. Lett. {\bf 60}, 1057 (1988); {\em ibid.} Pys. Rev. B {\bf
  39}, 2344 (1988).
\bibitem{Millis93} A. J.Millis, Phys. Rev.  B {\bf 48}, 7183 (1993).
%
%
\bibitem{CS93} A. V. Chubukov and S. Sachdev, Phys. Rev. Lett. {\bf
    71}, (1993);  A. V. Chubukov, S. Sachdev, and J. Ye, Phys. Rev. B {\bf 49},
11919 (1994).
%
\bibitem{SCS95} S. Sachdev, A. V. Chubukov, and A. Sokol, Phys. Rev. 
 B {\bf 51}, 14874 (1995).
%
\bibitem{MPchem}D. Pines and  P. Monthoux,  J . Phys.   Chem. of Solids,
{\bf 56}, 1651 (1995). 
%
\bibitem{SPS98} J. Schmalian, D. Pines, and B. Stojkovi\'c,
 Phys. Rev. Lett. {\bf 80}, 3839 (1998); {\em ibid} cond-mat/9804129 (unpublished).
%
\bibitem{2kf} We do not consider the case of an   instability driven by
$2{\bf k}_{\rm F}$ transitions, where the present expansion
 breaks down (see Ref.[7]).
%
\bibitem{SDW}A.P. Kampf and J.R. Schrieffer, Phys. Rev. B {\bf 42},
7967 (1990);  E. Dagotto, A. Nazarenko and A. Moreo, Phys. Rev.
Lett {\bf 74}, 310 (1995); M. Langer, J. Schmalian, S Grabowski,
and K. H. Bennemann, Phys. Rev. Lett. {\bf 75}, 4508 (1995);
 A.V. Chubukov and D.K. Morr, Phys.
 Rep., {\bf 288}, 355 (1997);  R. Preuss, W. Hanke, C. Gr\"ober, and H. G. Evertz,
Phys. Rev. Lett. {\bf  79} 1122 (1997). 
%
%
\bibitem{Schrieffer} J. R. Schrieffer, J. of Low Temp. Physics, {\bf 99}, 397 (1995).
%
%
%
\bibitem{VT97} Y. M  Vilk and A. M. S. Tremblay, J. Phys. I (France)  {\bf 7},
1309 (1997).
%
\bibitem{comment} 
A. V. Chubukov, Phys. Rev. B {\bf 52}, R3840 (1995);
B. L. Altshuler {\em et al.},
{\em ibid} 5563 (1995)). 
%
\bibitem{CS98} A. V. Chubukov, cond-mat/97101763 (unpublished); 
A. V. Chubukov and
J. Schmalian, Phys. Rev. B {\bf 57}, R11085 (1998). 
%
\bibitem{Adler}S. L. Adler, Phys. Rev. {\bf 137}, B1022 (1965);
D. V. Volkov {\em et al.}, Sov. Phys. Solid State {\bf 15}, 1396 (1971).
%
\bibitem{Subir94} S. Sachdev, Phys. Rev. B {\bf 49}, 6770 (1994).
%
\bibitem{ohg94} S. Ohsugi {\em et al.}, J. Phys. Soc. Jpn. {\bf 63}, 700 (1994).
\bibitem{Grilli} A. Gamba, M. Grilli, and C. Castellani (unpublished).
\end{thebibliography}
\end{document}